\documentclass[12pt,preprint]{aastex}

\begin{document}
\def\ch{{\bf changed}\ }


\title{A $1''$ Telescope: The Optimal Approach to Bright-Star Planetary Transits}

\author{Joshua Pepper, Andrew Gould and  D.\ L.\ DePoy}
\affil{Department of Astronomy, Ohio State University, Columbus, Ohio 43210}
\email{pepper@astronomy.ohio-state.edu, gould@astronomy.ohio-state.edu, 
depoy@astronomy.ohio-state.edu}

\begin{abstract}

Planetary transits of bright stars, $V<10$, offer the best opportunity
for detailed studies of extra-solar planets, such as are already being
carried out for HD209458b.  Since these stars are rare, they should
be searched over the entire sky.  In the limits of zero read-out time, 
zero sky noise, and perfect optics, the sensitivity of an all-sky
survey is independent of telescope aperture: for fixed detector size
and focal ratio, the greater light-gathering power of larger telescopes
is exactly cancelled by their reduced field of view.  Finite read-out
times strongly favor smaller telescopes because exposures are longer
so a smaller fraction of time is wasted on readout.  However, if the
aperture is too small, the sky noise in one pixel exceeds the stellar
flux and the field of view becomes so large that optical distortions
become unmanageable.  We find that the optimal aperture is about $1''$.
A one-year survey using such a $1''$ telescope could detect essentially
all hot-jupiter transits of $V<10$ stars observable from a given site.

\end{abstract}


\section{Introduction}

	In Pepper, Gould \& DePoy (astro-ph/0208042), we argued that all-sky
surveys are the best way to find planets transiting bright $V<10$ stars.
Such transits offer the best opporunity for detailed studies of planets.
We showed that the number of systems probed is given by
\begin{equation} \label{equfin}
N_{p} = \frac{4}{3} \pi n \eta \, d_{0}^{3} \, \left( \frac{R_{0}}{a_{0}} 
\right) \left( \frac{L}{L_{0}} 
\right)^{3/2} \left( \frac{R}{R_{0}} \right)^{-7/2} \left( \frac{a}{a_{0}}
\right)^{-5/2} \left( \frac{r}{r_{0}} \right)^{6} ,
\end{equation}
where $d_0$ is the maximum distance at which an $i = 90^{\rm o}$ transit can be
detected for a star of luminosity $L_{0}$, radius $R_{0}$, with a planet at 
semi-major axis $a_{0}$ and radius $r_{0}$, and where $n$ is the local number
density of such stars and $\eta=0.719$ is a numerical factor.

We then showed that the sensitivity of a given survey will depend almost
entirely on $\gamma$, the number of photons collected from a fiducial star 
of some fixed designated magnitude, which we arbitrarily chose to be $V=10$.
That is, it will not depend on the details of the all-sky observing program.
We then normalized equation (\ref{equfin}) in terms of $\gamma$, 
\begin{equation} \label{equfin2}
N_{p} = 600 F(M_{V}) \left( \frac{a}{a_{0}} \right)^{-5/2} \left( \frac{r}{r_{0}} 
\right)^{6} \left( \frac{\gamma}{\gamma_{0}} \right)^{3/2} \left( \frac{\Delta 
\chi_{\rm min}^{2}}{36} \right)^{-3/2}
\end{equation}
where we adopted $\gamma_{0} = 2 \times 10^{7}$, $a_{0}=20 \, R_{\odot}$, 
$r_{0} = 0.10 \, R_{\odot}$, and where we have made our evaluation at 
$M_{V} = 5$ (i.e. $R = 0.97 \, R_{\odot}$, $V_{\rm max} = 10$, 
$d_{0} = 100 \, $pc, and  $n = 0.004 \, {\rm pc}^{-3}$). Here,
$F(M_V)$ is a function, which is shown in Figure \ref{figF}, and 
$\Delta \chi^2_{\rm min}$ is the minimum $\chi^2$ improvement of
a transit-model relative to a constant-flux model, which is set to avoid
excessive false positives. Note that $\gamma_0=2 \times 10^{7}$ corresponds 
approximately to 1000 20-second exposures with a 5 cm telescope and a 
broadened $(V+R)$ type filter for one $V = 10$ mag fiducial star.

	Our goal was to apply this formula to the problem of telescope
design in a subsequent paper.  However, in light of the referee report,
we decided to append our work on telescope design as an additional chapter
of the original paper.  The following is that chapter.

\section{Implications for Telescope Design}

\begin{figure}[t]
\epsscale{1.0}
\plotone{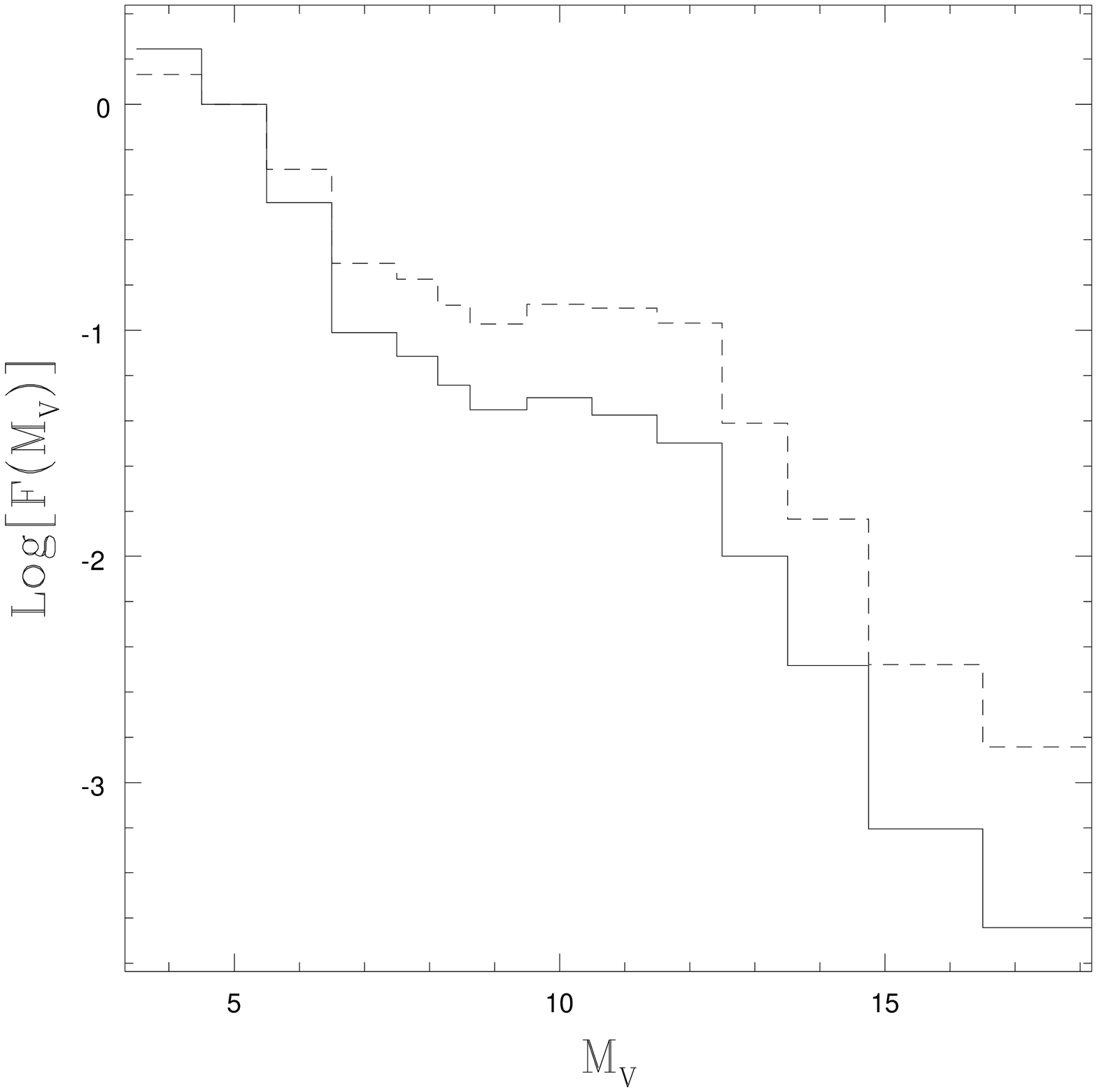}
\caption{The relative number of potential transiting systems $F(M_{V})$ probed for 
fixed planetary radius $r$ and semi-major axis $a$ as a function of $M_{V}$.  
The solid line applies to a uniform distribution of stars -- to model the 
immediate solar neighborhood.  The dashed line applies to a thin disk -- to 
model a search of a large portion of the Galactic disk.  The distributions are 
arbitrarily scaled such that $F(M_{V}=5)=1$.}
\label{figF}
\end{figure}

We now apply the general analysis of astro-ph/0208042
to the problem of optimizing telescope design for
quickly locating a ``large'' number of bright transiting systems.  Since
only one such system is now known, we define ``large'' as ${\cal O}(10)$.
 From equation (\ref{equfin2}) and the 0.75\% frequency of hot jupiters
measured from RV surveys, there are roughly five such systems to be discovered
over the whole sky per magnitude for $V_{\rm max}=10$.  Hence, from Figure 
\ref{figF}, of order 15 are to be discovered from all spectral types.  
It would,
of course, be possible to discover even more by going fainter, but setting 
this relatively bright 
limit is advisable for three reasons.  First, as we argued in
the introduction, the brightest transits are the most interesting 
scientifically, and most of the transits detected in any survey will be
close to the magnitude limit.  Second, as we discuss below, a wide range,
$\Delta V = V_{\rm max}-V_{\rm min}$, of survey sensitivity can only be
achieved at considerable cost to the observing efficiency.  Hence, if 
high efficiency is to be maintained, setting $V_{\rm max}$ fainter means 
eliminating the brightest (most interesting) systems.  Third, at 
$V_{\rm max}=10$, we are already reaching distances of 100 pc for G stars.
Hence the number of transits observed in fainter surveys will not continue
to grow as $d_0^3$ as in equation (\ref{equfin}).

In previous sections, we ignored the loss of sensitivity to systems
that are brighter than $V_{\rm min}$, which is set by saturation of the
detector (or more precisely, by the flux at which detector non-linearities
can no longer be accurately calibrated).  This fraction is $10^{-0.6\Delta V}$,
or 6\% for $\Delta V=2$, which we therefore adopt as a sensible goal.
That is, we wish to optimize the telescope design for,
\begin{equation}
8 = V_{\rm min} < V < V_{\rm max} = 10.
\label{equvminmax}
\end{equation}
(In any event, essentially all stars $V<8$ have already been surveyed
for XSPs using RV, and the problem of determining which among the 
planet-bearers have transits is trivial compared to the problem of conducting
an all-sky photometric variability survey.)

Optimization means maximizing the photon collection rate, $\gamma/T$, where
$T$ is the duration of the experiment and, again, $\gamma$ is the total
number of photons collected from a fiducial $V=10$ mag star.  Explicitly,
\begin{equation} \label{equGam1}
\gamma = K {\cal E} T D^2 \frac{(\Delta \theta)^2}{4\pi},
\end{equation}
where $\Delta\theta$ is the angular size of the detector,  $D$ is the 
diameter of the primary-optic, ${\cal E}$ is the fraction of the time
actually spent exposing, and $K$ is a constant that depends on the telescope,
filter, and detector throughput.  For our calculations, we assume
$K= K_0\equiv 40 \, e^-\,\rm cm^{-2}\,s$, which is appropriate for a broad 
$(V+R)$ filter
and the fiducial $V=10$ mag star.  The design problems are brought into sharper
relief by noting that $\Delta\theta = {\cal L}/D{\cal F}$, where
${\cal L}$ is the linear size of the detector and ${\cal F}$ is the focal
ratio, or $f$/\#, of the optics.  Equation (\ref{equGam1}) then becomes
\begin{equation} 
\gamma = {K {\cal E} {\cal L}^2 T \over 4\pi{\cal F}^2}.
\label{equGam2}
\end{equation}

That is, almost regardless of other characteristics of the system, the 
camera should be made as fast as possible.  We will adopt ${\cal F}=2$, beyond
which it becomes substantially more difficult to design the optics.
A more remarkable feature of equation (\ref{equGam2}) is that all explicit 
dependence on the size of the primary optic has vanished: a 1'' telescope
and an 8m telescope would appear equally good!  Actually, as we now show,
there is a hidden dependence of ${\cal E}$ on $D$, which favors small 
telescopes.

The global efficiency ${\cal E}$ can be broken down into two factors,
${\cal E} = {\cal E}_0 {\cal E}_S$, where ${\cal E}_0$ is the fraction of
time available for observing (i.e., during which the sky is dark, the weather
is good, etc.), and ${\cal E}_S$ is the fraction of this available observing
time that the shutter is actually open.  The first factor is not affected
by telescope design and so will be ignored for the moment.  The second factor
should be maximized.  Since the readout time is fixed, the smaller is the
telescope aperture, the longer can be the exposures before a $V_{\rm min}=8$
mag star saturates, and so the smaller fraction of time is lost to read-out.
For large-format detectors, the pixel size is typically $\Delta x_p=15\,\mu$m,
for which significant non-linearities set in at about $60,000\,e^-$.
Hence, exposure times are 
\begin{equation}
T_{\rm exp} = 38\,{\rm s}\,{K_0\over K}
\biggl({D\over 2.5\,\rm cm}\biggr)^{-2}
\label{equtexp}
\end{equation}
That is, the exposure time for a $1''$ ``telescope'' is already of
the order of typical read-out times for large-format detectors.
Clearly, smaller is better, but are there constraints from going too small?

One potential constraint comes from sky noise.  To stay within the photon-noise
limited regime, which has been assumed in all of our calculations, the
sky inside one pixel should be at least one magnitude fainter than 
$V_{\rm max}$, and preferably two mags fainter.  Assuming the sky brightness
in our
broadening passband reaches a maximum of $V=19.7\,\rm arcsec^{-2}$, and again 
assuming $\Delta x_p=15\,\mu$m pixels, the sky in one pixel is
\begin{equation}
V_{\rm sky} = 10.8 + 5\log\biggl(
{D\over 2.5\,\rm cm}\,{{\cal F}\over 2}
\biggr).
\label{equsky}
\end{equation}
Hence, the sky is a bit bright for a $1''$ telescope, but appears quite
satisfactory for a $2''$.

Finally, one must be careful that the field of view is not too large, or
the focal-plane distortions at its edges will be difficult (and expensive)
to correct.  For example, for a $4k\times 4k$ detector with 
$\Delta x_p=15\,\mu$m pixels, 
$\Delta \theta = 68^\circ (D/2.5\,{\rm cm})^{-1}({\cal F}/2)^{-1}$.  
At $1''$, this
is probably too large to correct at reasonable expense.  With this size 
detector, both sky-noise considerations, and problems in optics design 
argue for a $1.5''$ telescope.  
However, from equation (\ref{equtexp}),
such a ``large'' telescope will most likely be dominated by read-out time.
In summary, we conclude that a $1''$ to $2''$ telescope equipped with a 
$4x\times 4k$, ${\cal L}=6\,$cm detector and with ${\cal F}=2$ focal ratio
is optimal for this observing problem.   
Among all existing transit programs of which we are aware, the WASP
telescope ($D=2.5''$ lens, ${\cal F}=2.8$ focal ratio,
$2k \times 2k$, ${\cal L}=3\,$cm detector, \citealt{str02}) 
comes closest to meeting these design specifications.

Adopting $\gamma=\gamma_0$, $K=K_0$, ${\cal E}_{0}=20\%$, 
${\cal E}_{S}=50\%$, ${\cal L}=6\,$cm, and ${\cal F}=2$, we find from
equation (\ref{equGam2}) that the required duration of the experiment using
our optimally-designed telescope is
$T=3\,\rm months$.  Clearly, roughly a year is required just to get around
the sky.  The S/N acquired during such a year-long search would therefore
roughly double the minimum requirements calculated here.

\acknowledgments This work was supported in part by
grant AST 02-01266 from the NSF.

\end{document}